\documentclass{PoS}

\title{Convergence issues in ChPT: a lattice perspective}

\ShortTitle{Convergence issues in ChPT: a lattice perspective}

\author{\speaker{Stephan D\"urr}%\thanks{A footnote may follow.}
        \\
        Wuppertal University and IAS/JSC Forschungszentrum J\"ulich\\
        E-mail: \email{durr\,(AT)\,itp.unibe.ch}}

\abstract{This review addresses the practical convergence of the ChPT series
in the p-regime. In the SU(2) framework there is a number of new results, and
improved estimates of $\bar\ell_3$ and $\bar\ell_4$ are available. In the SU(3)
framework few new lattice computations have appeared and the improvement in the
precision of the low-energy constants $L_i$ is comparatively slow. I sketch
some of the convergence issues genuine to extensions of ChPT
which include additional sources of chiral symmetry breaking (finite lattice
spacing) and/or violations of unitarity (different sea and valence quark
masses). Finally, it is pointed out that the quark mass ratios
$m_u/m_d$, $m_s/m_d$ happen to be such that no reordering of the chiral
series is needed to accommodate the experimental pion and kaon masses.}

\FullConference{2013 Kaon Physics International Conference\\
                Apr 29-30 and May 1st\\
                University of Michigan, Ann Arbor, Michigan - USA}

\newcommand{\be}{\beta}

\newcommand{\ch}{\chi}

\newcommand{\bdm}{\begin{displaymath}}
\newcommand{\edm}{\end{displaymath}}
\newcommand{\bea}{\begin{eqnarray}}
\newcommand{\eea}{\end{eqnarray}}
\newcommand{\beq}{\begin{equation}}
\newcommand{\eeq}{\end{equation}}

\newcommand{\mr}{\mathrm}

\newcommand{\Nf}{N_{\!f}}

\newcommand{\MeV}{\,\mr{MeV}}
\newcommand{\GeV}{\,\mr{GeV}}

\newcommand{\fm}{\,\mr{fm}}

\newcommand{\MSbar}{{\overline{\mr{MS}}}}

\newcommand{\Mpi}{M_\pi}
\newcommand{\Fpi}{F_\pi}
\newcommand{\Mka}{M_K}

\newcommand{\fpi}{f_\pi}

\begin{document}

%%%%%%%%%%%%%%%%%%%%%%%%%%%%%%%%%%%%%%%%%%%%%%%%%%%%%%%%%%%%%%%%%%%%%%%%%%%%%%%%
%%% The deadline for submission has been set to May 31st, 2013.
%%%%%%%%%%%%%%%%%%%%%%%%%%%%%%%%%%%%%%%%%%%%%%%%%%%%%%%%%%%%%%%%%%%%%%%%%%%%%%%%
%%% Contributions should not exceed 10 pages unless you feel that the quality
%%% of your submission will really be suffering from this limitation.
%%%%%%%%%%%%%%%%%%%%%%%%%%%%%%%%%%%%%%%%%%%%%%%%%%%%%%%%%%%%%%%%%%%%%%%%%%%%%%%%

\section{Introduction}

\begin{figure}[tb]
\includegraphics[height=3.9cm]{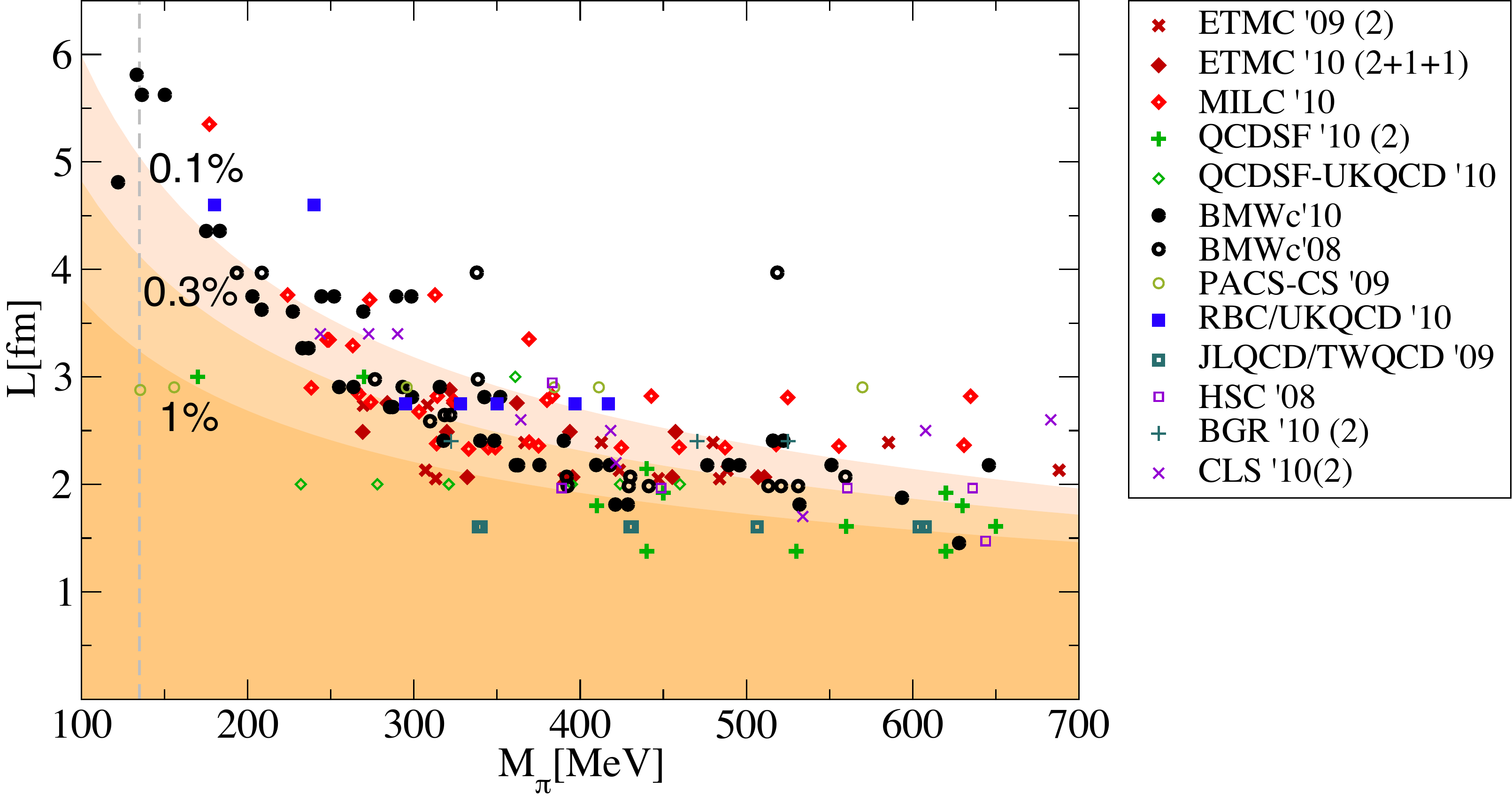}%
\includegraphics[height=3.9cm]{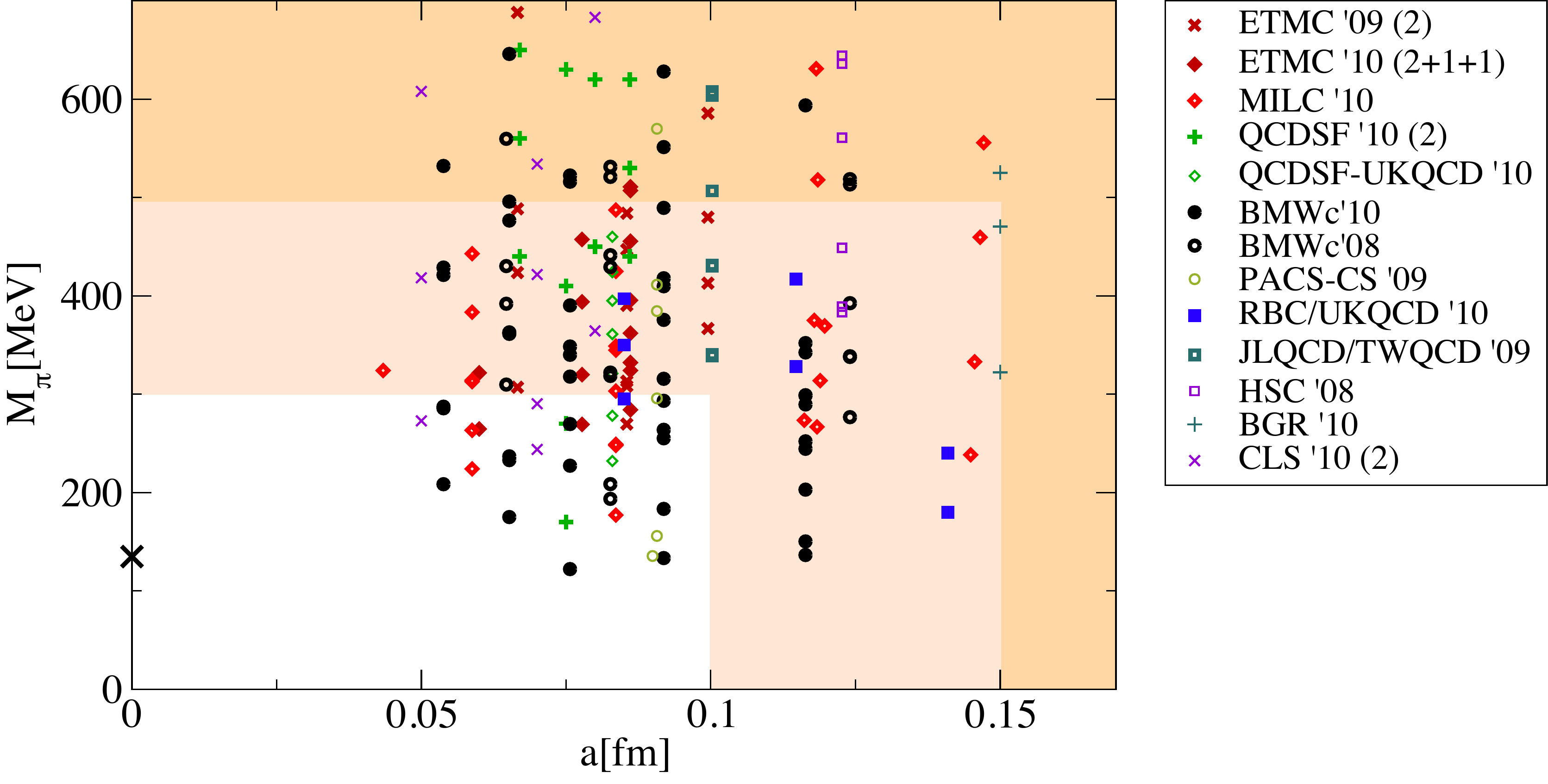}%
\caption{\label{fig1}
Overview of the datasets accumulated in QCD ($\Nf\geq2$) by various
collaborations as of 2011, plotted in the $L$ versus $\Mpi$ or the $\Mpi$
versus $a$ plane. The shaded backgrounds indicate the size of systematic
effects due to the finite spatial volume or the chiral and continuum
extrapolation. Figure taken from Ref.\,\cite{Fodor:2012gf}.}
\end{figure}

Over the past few years computations in lattice QCD have greatly progressed.
Today we aim for simulating $\Nf=2+1$ QCD (i.e.\ with a degenerate up and
down quark mass $m_{ud}$ and a separate strange quark mass in the determinant)
right at the physical mass point $m_{ud}=(m_u^\mr{phys}+m_d^\mr{phys})/2$ and
$m_s=m_s^\mr{phys}$ where $\Mpi^2$ and $2\Mka^2-\Mpi^2$ take their physical
values, in large boxes (up to $6\fm$ to control finite-size effects) and at
several lattice spacings $a$ (to allow for a continuum extrapolation $a\to0$).
This goal has been reached by the Wuppertal-Budapest collaboration
(staggered fermions), the BMW collaboration (Wilson fermions), the PACS-CS
collaboration (ditto), the MILC collaboration (staggered fermions) and the
RBC/UKQCD collaboration (domain-wall fermions) -- see the talk by Bob Mawhinney
\cite{Mawhinney} for more details and Fig.\,\ref{fig1} for an illustration
(as of 2011).

These developments have a strong impact on the relation between Lattice QCD
(LQCD) and Chiral Perturbation Theory (ChPT).
In the past ChPT was used to guide the ``chiral extrapolation'' by which
lattice physicists meant the extrapolation to $\Mpi\simeq135\MeV$.
In addition ChPT proved useful to correct data for the impact of the finite
spatial box-size $L$, e.g.\ by providing the factor $M_X(\infty)/M_X(L)$ to be
applied on the numerical data $M_X(L)$ for the mass of the state $X$.
Now, the former application is less relevant, while the latter one is still
extremely helpful (provided $L$ is large enough so that ChPT can be applied).
However, with todays lattices one can map out the quark mass dependence of
various observables, and this provides a unique opportunity to determine
the low-energy constants (LECs) of ChPT.
The only ``caveat'' is that one must make sure that the data are in a regime
where ChPT can be applied, i.e.\ converges (in a practical sense) well.
The goal of this review is to provide examples of ``good'' and ``bad''
convergence and to discuss the status of lattice determinations of LECs in the
SU(2) and SU(3) chiral frameworks.

%%%%%%%%%%%%%%%%%%%%%%%%%%%%%%%%%%%%%%%%%%%%%%%%%%%%%%%%%%%%%%%%%%%%%%%%%%%%%%%%

\section{Some Lattice and ChPT terminology}

The purpose of this section is to recall some Lattice and ChPT terminology; the
reader familiar with these is invited to move directly to Sec.\,\ref{sec:SU2}.

\begin{figure}[tb]
\centering
\vspace*{-2mm}
\includegraphics[width=6.6cm]{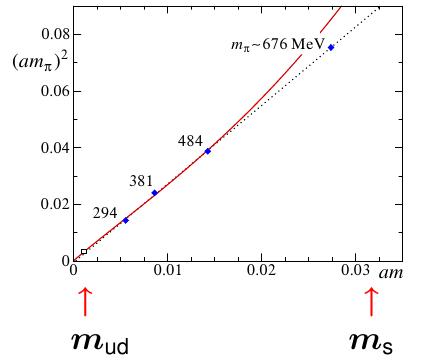}%
\vspace*{-2mm}
\caption{\label{fig2}
An early result for $\Mpi^2$ versus $m_q$ in $\Nf=2$ QCD. The data are
consistent with a linear behavior, yet ChPT predicts a curvature from which one
is supposed to extract $\bar\ell_3$. Figure taken from
Ref.\,\cite{Luscher:2005mv}.}
\end{figure}

ChPT is a rigorous framework to compute Green's functions of QCD, based on
$(i)$~symmetry, $(ii)$~analyticity and $(iii)$~unitarity.
It is organized as an expansion in external momenta $p^2$ and quark masses
$m_q$.
At each order there is a number of new LECs which help govern the momentum and
quark-mass dependence of the Green's functions [at LO there are 2 parameters
$B$, $F$ in the SU(2) framework or $B_0$, $F_0$ in the SU(3) framework; at NLO
there are 7 parameters $\bar\ell_i$ for SU(2) or 10 parameters
$L_i^\mr{ren}(\mu)$ for SU(3)].
Those linear combinations of LECs which parameterize the $p$-dependence are
usually best determined in experiment.
By contrast, those linear combinations which determine the $m_q$-dependence are
hard to get from experiment (in nature the quark masses can be varied in
discrete steps only) and this creates an obvious opportunity for the lattice.

\begin{figure}[tb]
\centering
\vspace*{-6mm}
\includegraphics[width=10cm]{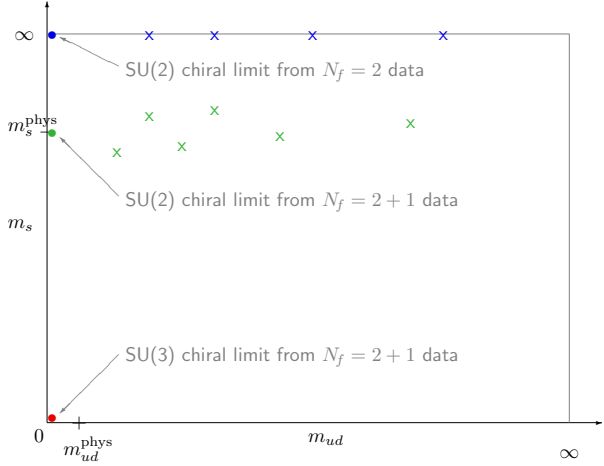}%
\vspace*{-10mm}
\caption{\label{fig3}
Cartoon of different data taking strategies in the $(m_{ud},m_s)$ plane.
Simulations of QCD with $\Nf=2$ work effectively at $m_s=\infty$.
Simulations with $\Nf=2+1$ tend to have $m_s$ values in the
vicinity of $m_s^\mr{phys}$; for a controlled extrapolation to the SU(3)
chiral limit additional data with $m_s\ll m_s^\mr{phys}$ are mandatory.}
\end{figure}

The standard counting rule is $p^2\sim m$, but early on it was difficult to
prove that the condensate parameter $B$ or $B_0$ is large enough to warrant
this counting (in phenomenology only the combination $Bm_q$ or $B_0m_q$ can be
determined).
Fig.\,\ref{fig2} displays a historical plot by L\"uscher \cite{Luscher:2005mv}
which shows that the lattice did step in to fill this gap: $\Mpi^2$ is in
remarkably good approximation linear in $m_q$, and the slope is just
$2B=2\Sigma/F^2$.
Moreover, the tiny deviation from linearity (which is not statistically
significant in these data) bears the knowledge of $\bar\ell_3$.
This illustrates that there is an enormous \emph{hierarchy of difficulty}
between determining the LECs at LO versus at NLO\,!

A few words on the relationship between $\Nf=2$ or $\Nf=2+1$ data and the
SU(2) or SU(3) chiral frameworks are in order.
Data with two degenerate dynamical flavors ($\Nf=2$) can only be analyzed with
SU(2) formulas.
The resulting LECs are logically different from those in nature, since the
latter bear an implicit knowledge of $m_s^\mr{phys}$ (and heavier flavors).
Also data with two degenerate light and a separate heavier flavor in the
determinant ($\Nf=2+1$) may be analyzed in the SU(2) framework,
cf.\ Fig.\,\ref{fig3}.
If $m_s$ was fixed at or near $m_s^\mr{phys}$ the resulting LECs may be
identified with the phenomenological SU(2) LECs, since the implicit dependence
of the latter on $m_c^\mr{phys}$ (and heavier flavors) is tiny.
In addition, $\Nf=2+1$ data may be analyzed in the SU(3) framework, if the
largest $m_s$ used is small enough to warrant the chiral expansion.
Hence, by increasing $m_s^\mr{max}$ the lattice may determine whether
``catastrophic failure'' occurs before or after reaching $m_s^\mr{phys}$.

Sometimes lattice physicists analyze their data with \emph{extended versions}
of ChPT which are designed to parameterize the effects of \emph{unitarity
violation} (which come from $m_q^\mr{sea}\neq m_q^\mr{val}$ a.k.a.\ ``partial
quenching'') and/or \emph{finite lattice spacing} (specific to the lattice
action used).
It is important to keep in mind that these new capabilities bring in new
convergence issues; it is well conceivable that there is a bound on the range
of $|m_q^\mr{sea}-m_q^\mr{val}|$ that these theories may describe.

%%%%%%%%%%%%%%%%%%%%%%%%%%%%%%%%%%%%%%%%%%%%%%%%%%%%%%%%%%%%%%%%%%%%%%%%%%%%%%%%

\section{Success with the chiral SU(2) framework \label{sec:SU2}}

\begin{figure}[tb]
\centering
\vspace*{-4mm}
\includegraphics[width=0.5\textwidth]{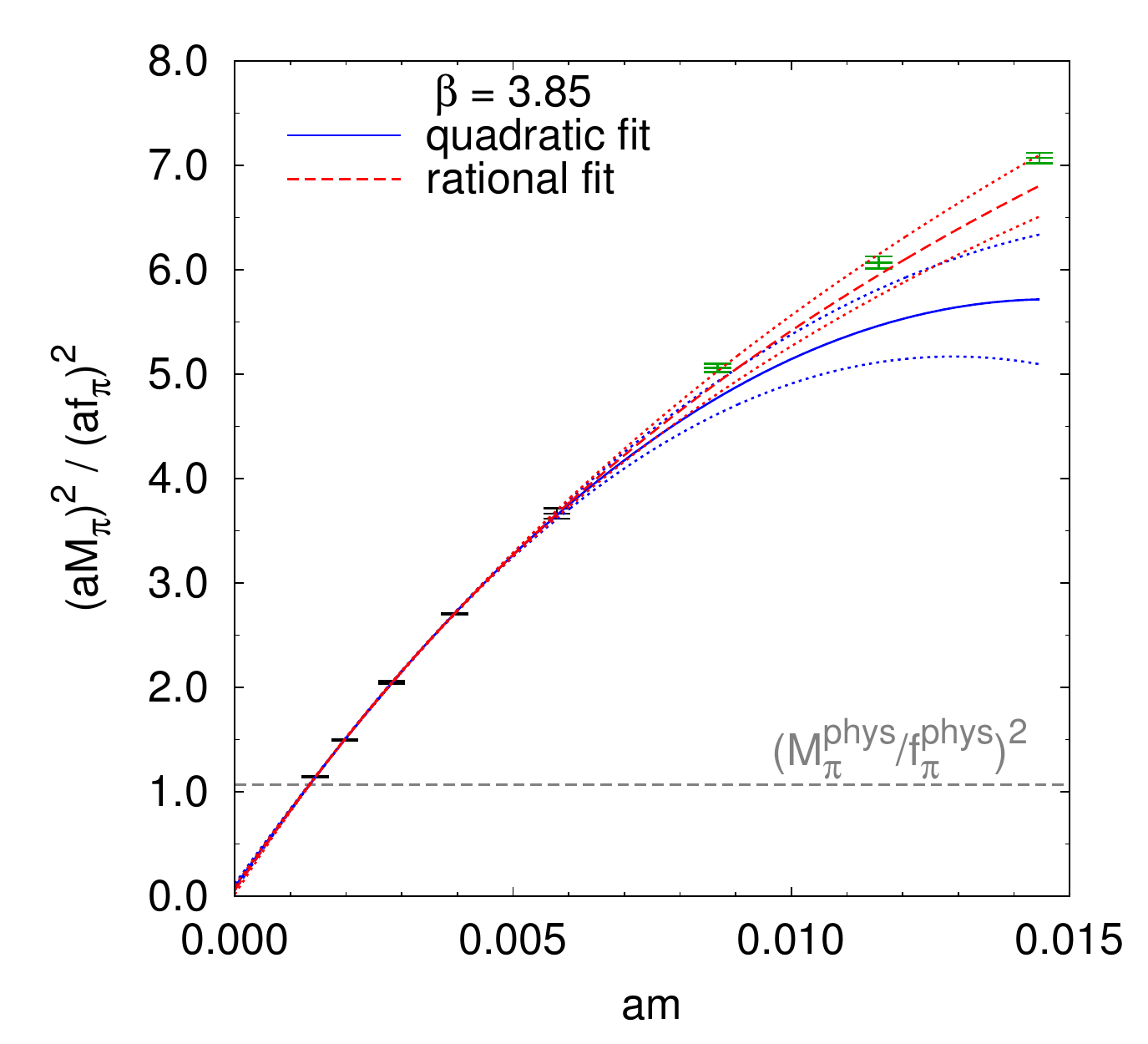}%
\includegraphics[width=0.5\textwidth]{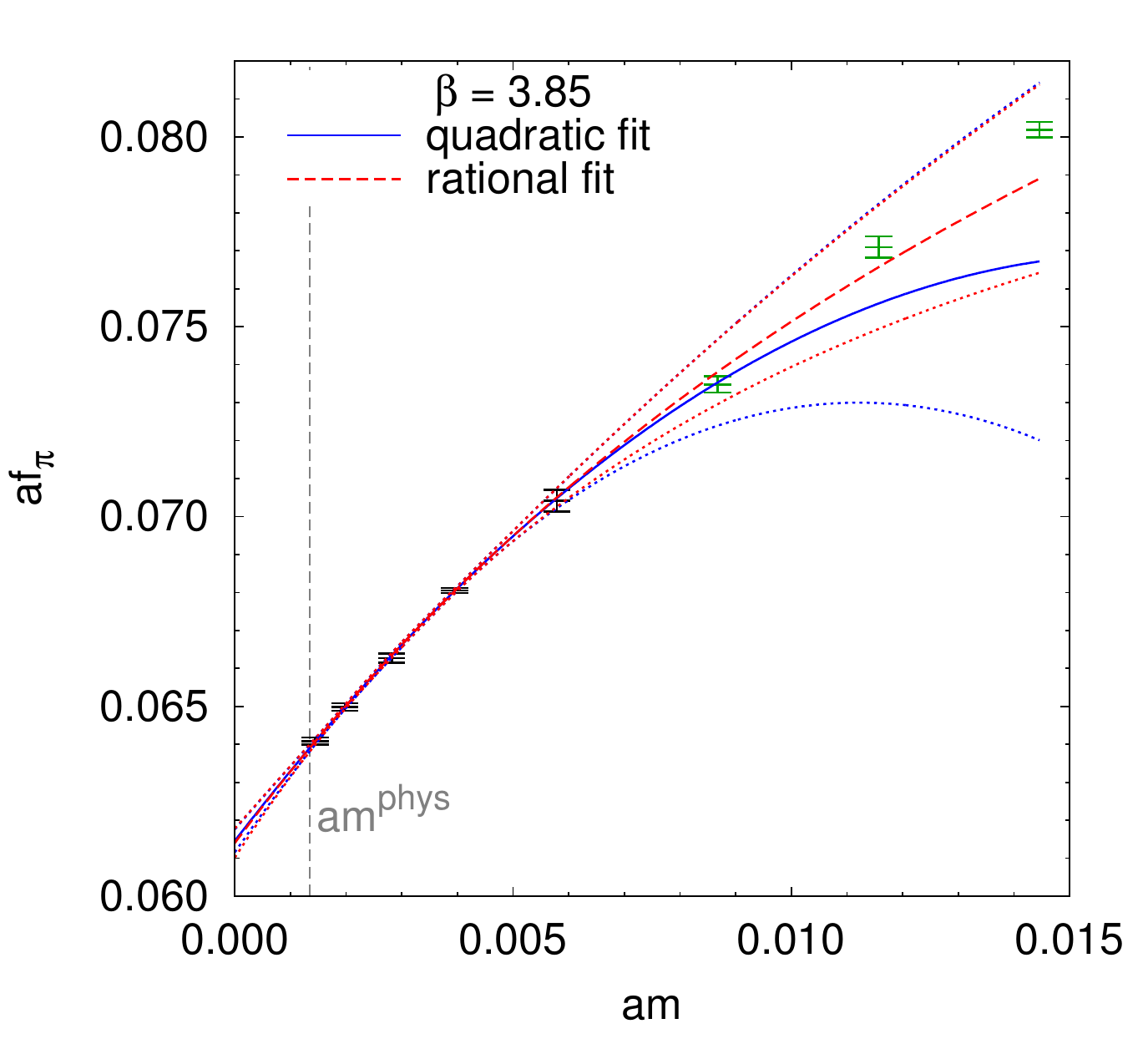}%
\vspace*{-2mm}
\caption{\label{fig4}
Strategy of Ref.\,\cite{Borsanyi:2012zv} for setting the scale and for
adjusting $m_{ud}$ to $m_{ud}^\mr{phys}$. For each lattice spacing (bare
coupling $\be$) the data for $(a\Mpi)^2/(a\fpi)^2$ are
interpolated/extrapolated to the point where this ratio assumes its physical
value. The respective quark mass in lattice units is $am_{ud}^\mr{phys}$
(left), and by comparing the respective $a\fpi$ to $\fpi^\mr{phys}$ one finds
$a$ (right). See text for details. Figure taken from
Ref.\,\cite{Borsanyi:2012zv}.}
\end{figure}

%In my opinion the first paper in which the lattice demonstrated its capability
%to investigate convergence issues in the SU(2) framework and to pin down the
%corresponding LECs with good control over the chiral systematics is
%Ref.\,\cite{Noaki:2008iy} by the JLQCD/TWQCD collaboration.
An early (and I think particularly nice) paper in which the lattice demonstrated
its ability to investigate convergence issues in the SU(2) framework and to pin
down the corresponding LECs with good control over the chiral systematics is
Ref.\,\cite{Noaki:2008iy} by the JLQCD/TWQCD collaboration.

A more recent paper which I would like to discuss in some detail (perhaps
because I'm an author) is \cite{Borsanyi:2012zv}.
It uses staggered $\Nf=2+1$ simulations with $m_s$ tuned to $m_s^\mr{phys}$ and
controls all sources of systematic error, including finite-size effects and
cut-off effects (besides the chiral range).
The scale is set by identifying the pion decay constant $\fpi=\sqrt{2}\Fpi$ at
the physical mass point with the PDG value, see Fig.\,\ref{fig4} for details
(there are some encouraging signs that the MILC collaboration might adopt this
simple and compelling scale-setting strategy in future works, too).

\begin{figure}[tb]
\centering
\vspace*{-4mm}
\includegraphics[width=0.5\textwidth]{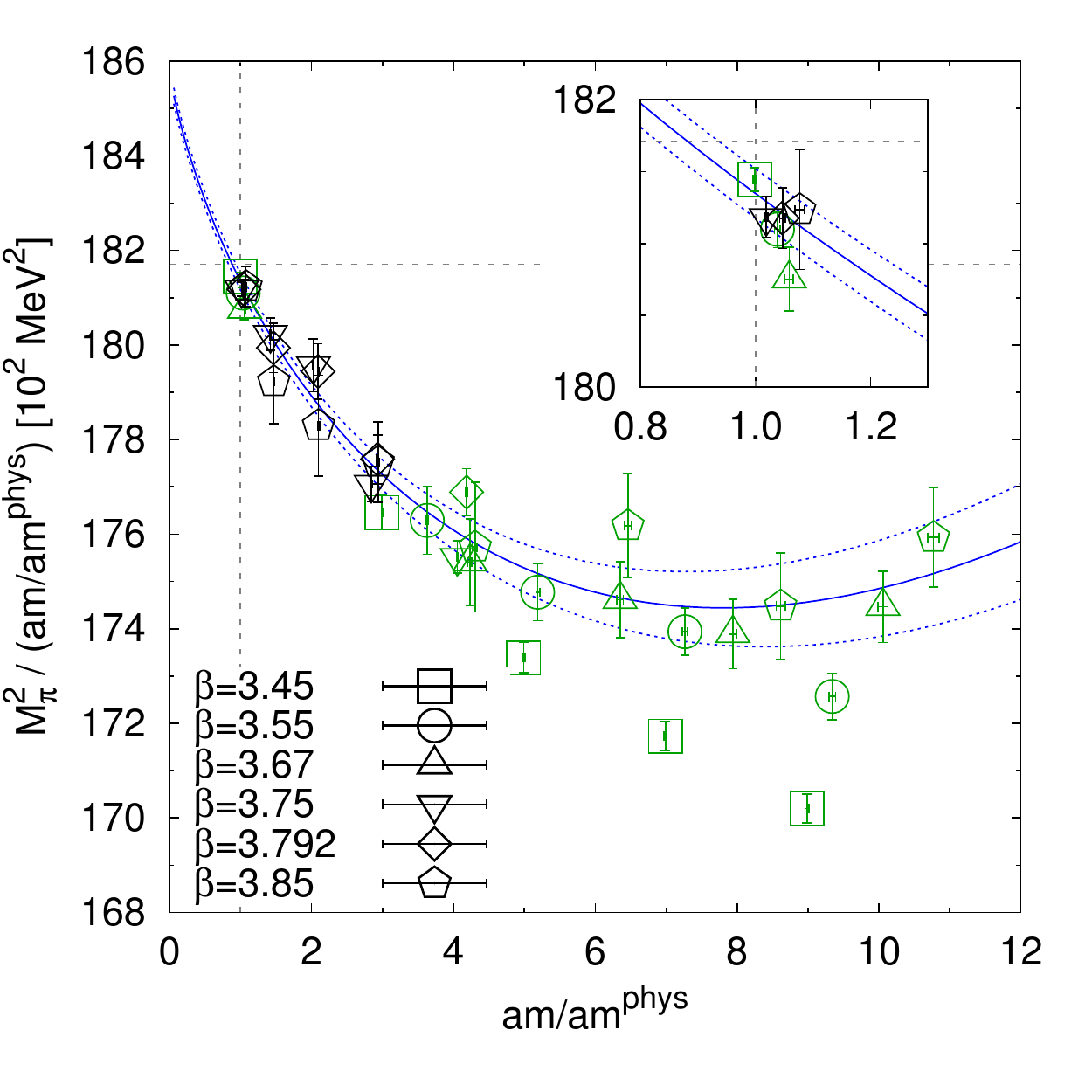}%
\includegraphics[width=0.5\textwidth]{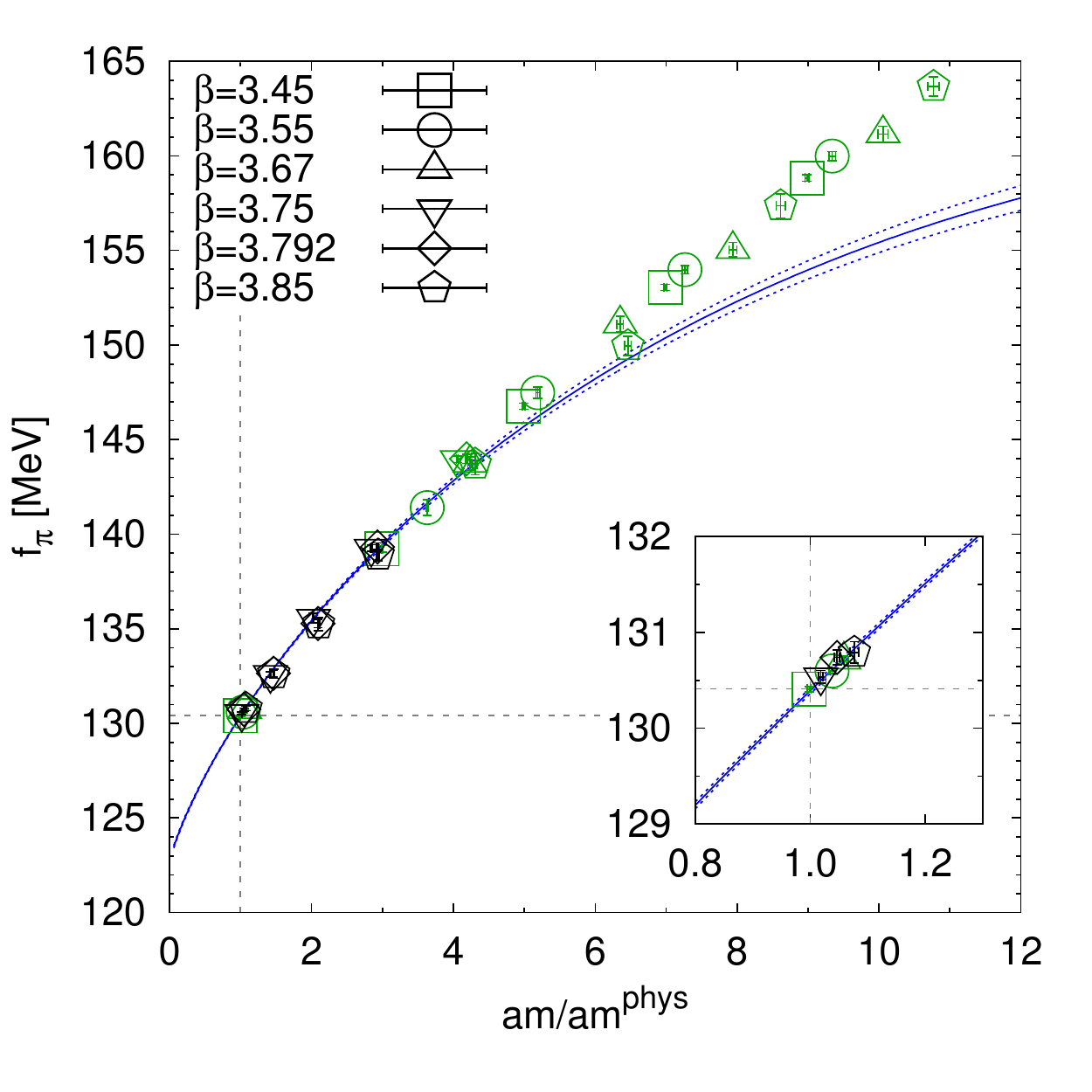}%
\vspace*{-4mm}
\caption{\label{fig5}
Plot of $\Mpi^2/(am_{ud})\cdot(am_{ud}^\mr{phys})$ versus
$(am_{ud})/(am_{ud}^\mr{phys})$ [left] and $\fpi\,[\mr{MeV}]$ versus
$(am_{ud})/(am_{ud}^\mr{phys})$ [right].The latter quantity has no cut-off
effects at the physical mass point, whereas the former one has
cut-off effects at the few-permille level (see inserts).
The LO+NLO fit includes data from the three finest lattices in the range
$135\MeV\leq\Mpi\leq240\MeV$ (black); other data (green) are disregarded.
Figure taken from Ref.\,\cite{Borsanyi:2012zv}.}
\end{figure}

\begin{figure}[tb]
\centering
\vspace*{-4mm}
\includegraphics[width=0.5\textwidth]{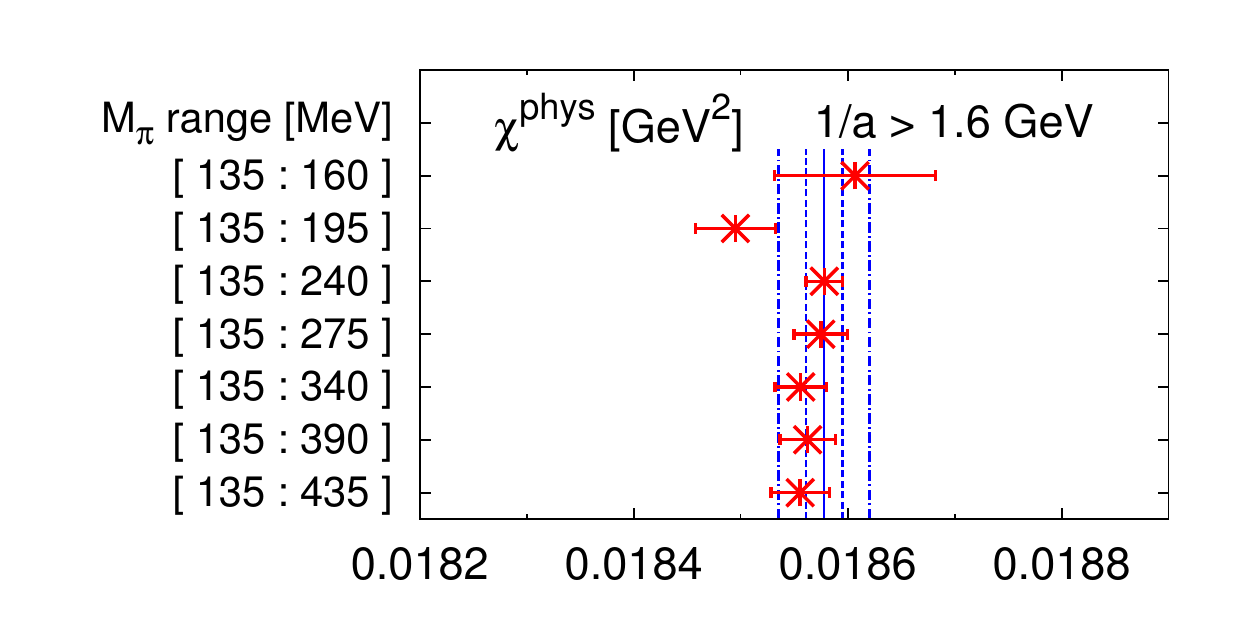}%
\includegraphics[width=0.5\textwidth]{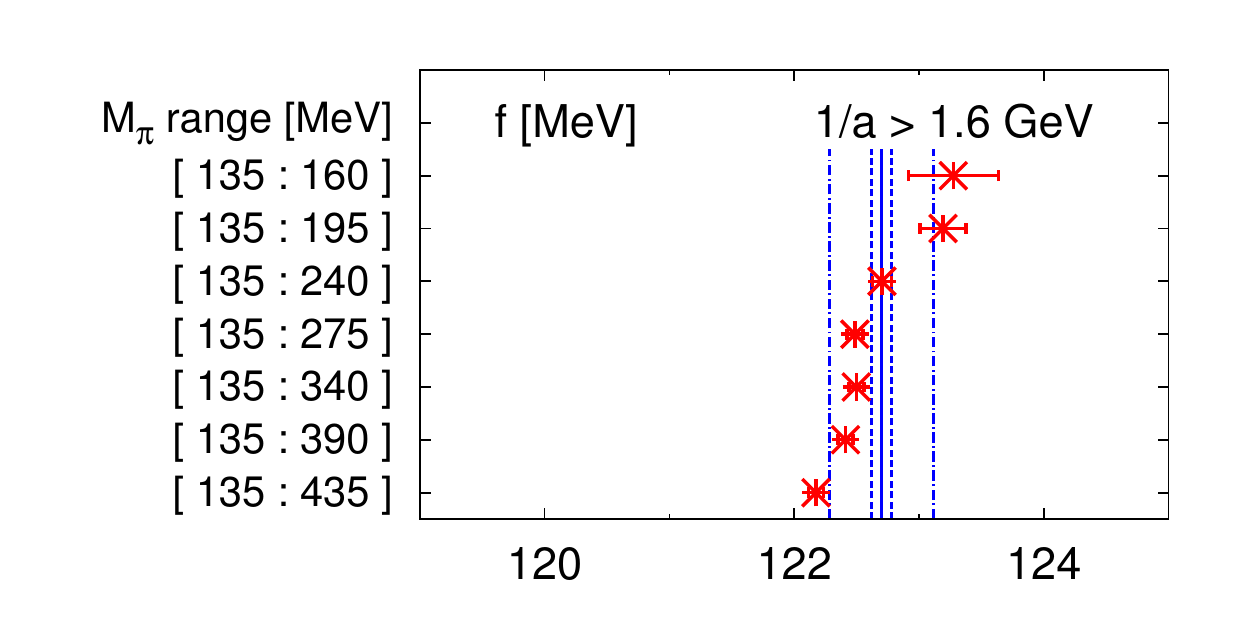}%
\\[-4mm]
\includegraphics[width=0.5\textwidth]{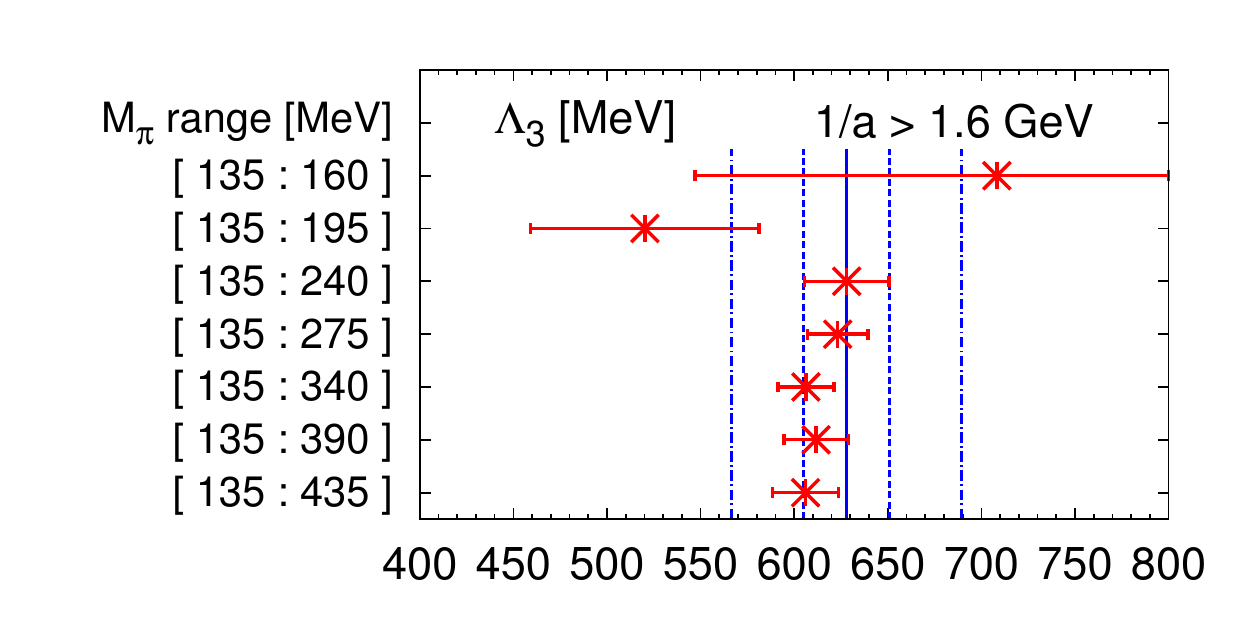}%
\includegraphics[width=0.5\textwidth]{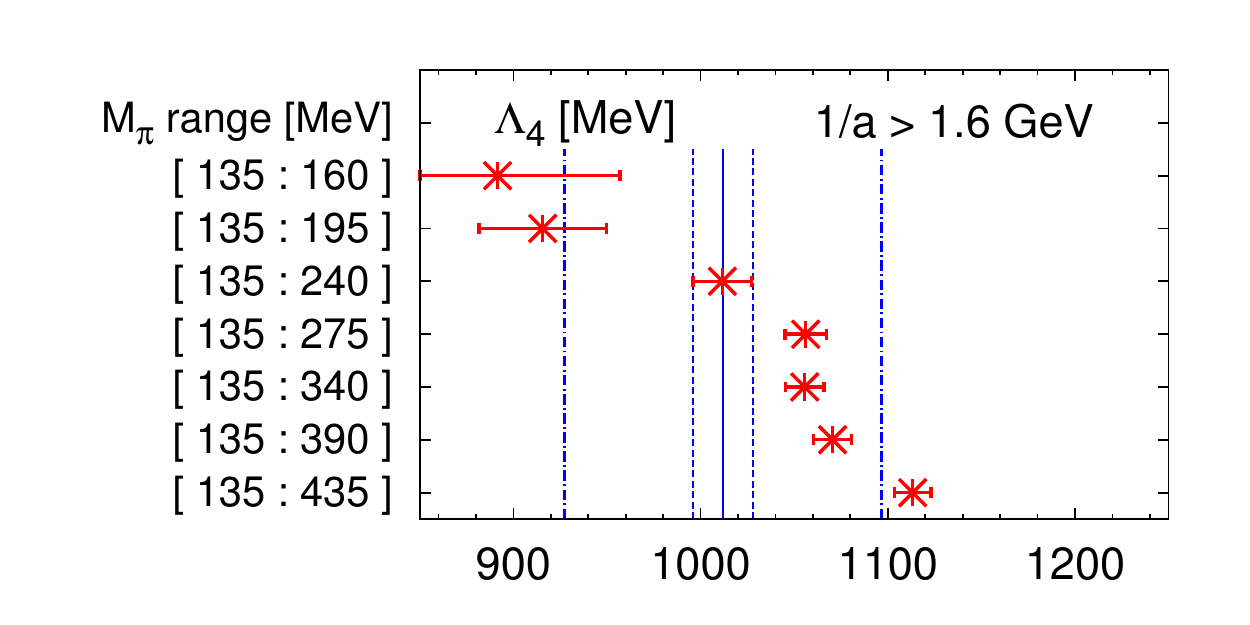}%
\vspace*{-2mm}
\caption{\label{fig6}
Behavior of the LECS at LO ($\ch=2Bm$ and $f\,[\mr{MeV}]$, top) and at NLO
($\Lambda_3\,[\mr{MeV}]$ and $\Lambda_4\,[\mr{MeV}]$, bottom) as a function of
the chiral range. Our preferred fit uses the range $135\MeV\leq\Mpi\leq240\MeV$;
the systematic error of the final result follows from the width of the
distribution. Figure taken from Ref.\,\cite{Borsanyi:2012zv}.}
\end{figure}

The LECs are determined by a joint fit of the standard LO+NLO SU(2) formulas
for $\Mpi^2/m_{ud}$ and $\Fpi$ as a function of $m_{ud}$ (the abscissa value 1
in Fig.\,\ref{fig5} indicates the physical pion mass).
We get a decent description of the data if we restrict the fit to
the three finest lattices (i.e.\ $a<0.13\fm$) and the mass range
$135\MeV\leq\Mpi\leq240\MeV$.
Alternative fit ranges affect $\chi^\mr{phys}=2Bm_{ud}^\mr{phys}$ and
$\Lambda_3$ (both extracted from $\Mpi^2/m_{ud}$ versus $m_{ud}$) less severely
than $f=\lim_{m_{ud}\to0}\fpi$ and $\Lambda_4$ (both extracted from $\fpi$ as a
function of $m_{ud}$), see Fig.\,\ref{fig6}.
The systematic uncertainty of the LECs is extracted from the variance over
the 7 chiral fit ranges (all other uncertainties are massively subdominant).

\begin{figure}[tb]
\centering
\vspace*{-4mm}
\includegraphics[width=0.46\textwidth]{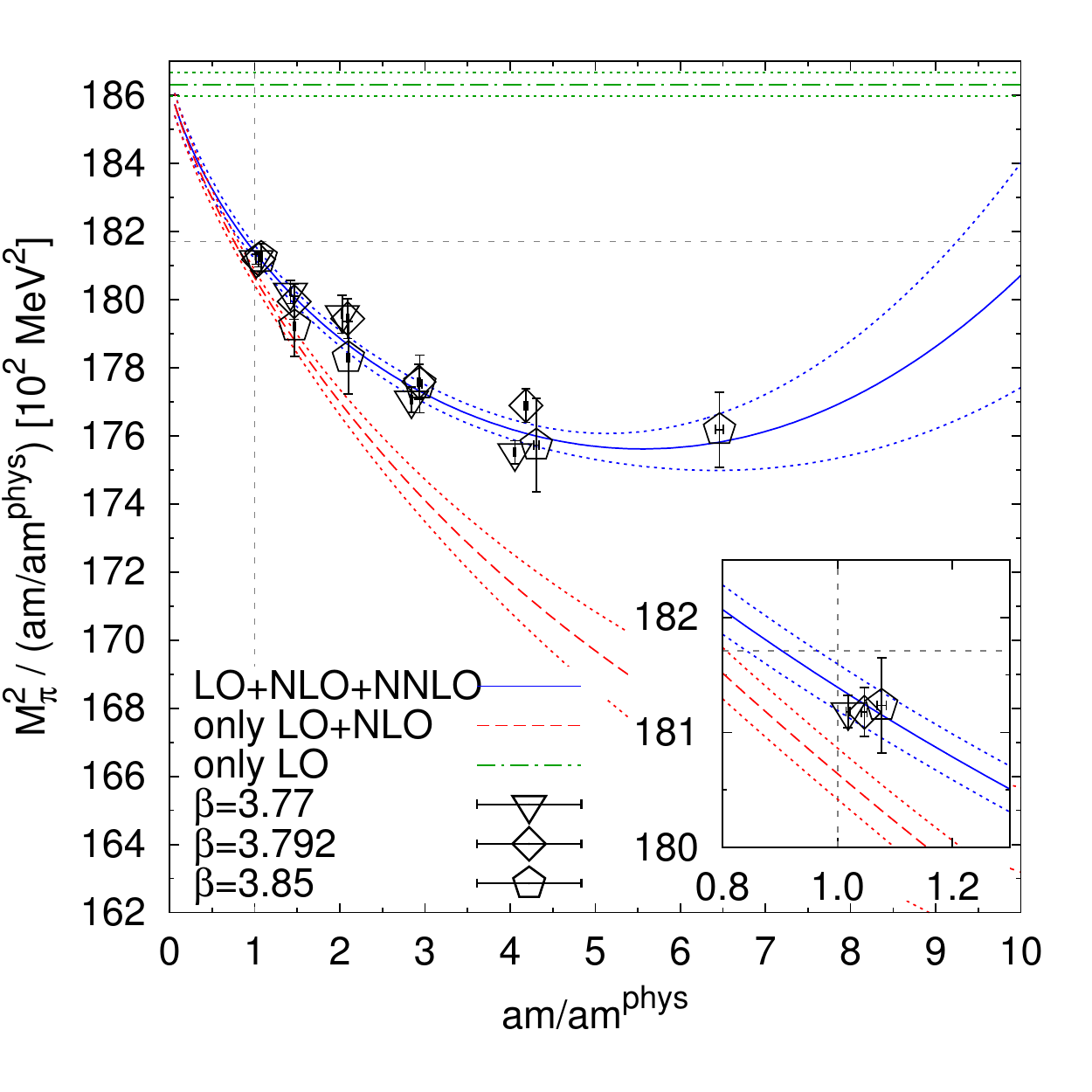}%
\includegraphics[width=0.46\textwidth]{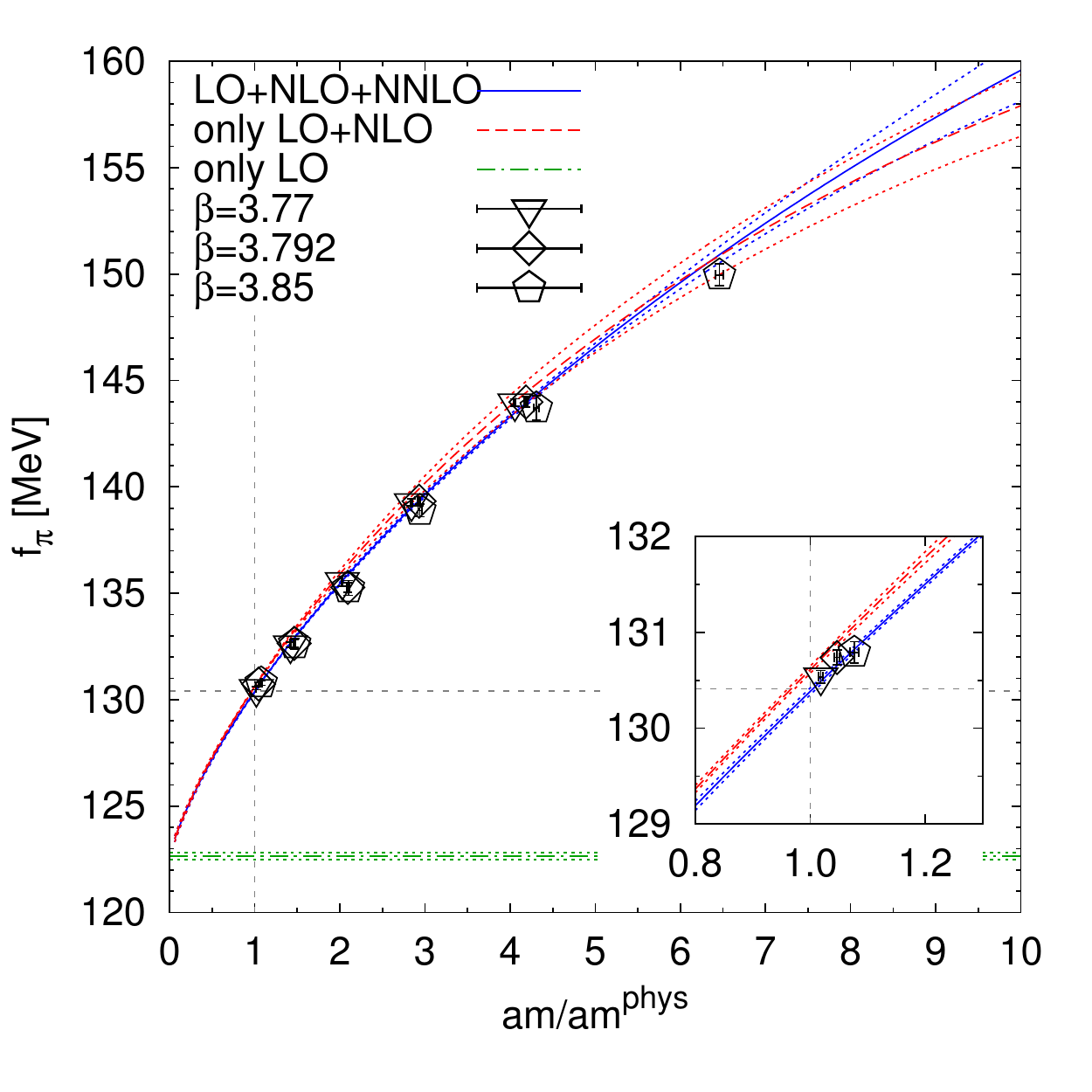}%
\\[-4mm]
\includegraphics[width=0.47\textwidth]{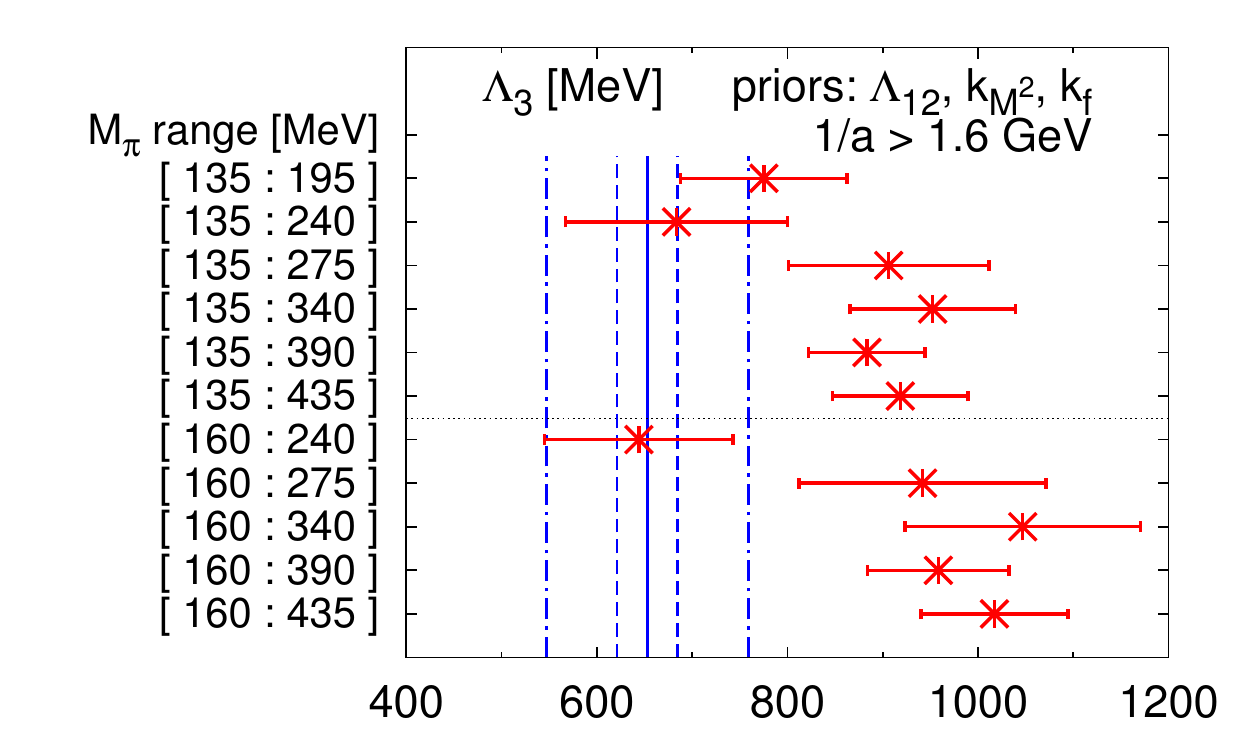}%
\includegraphics[width=0.47\textwidth]{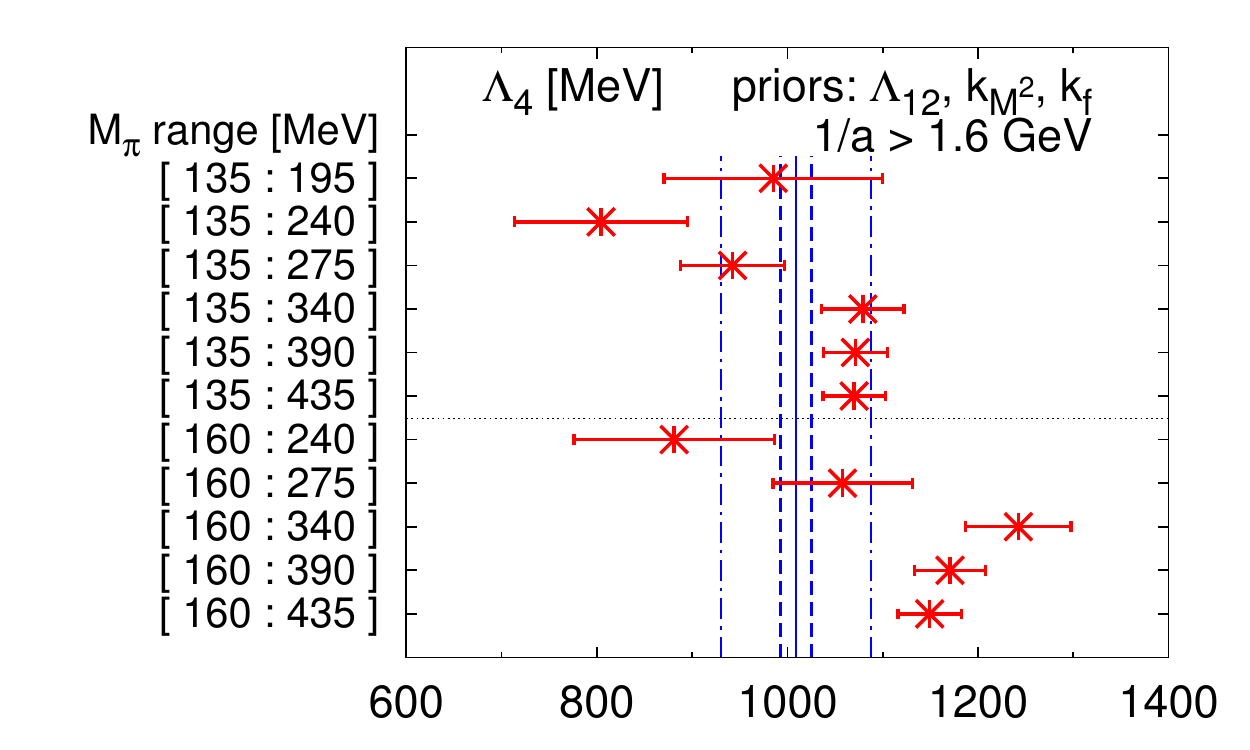}%
\vspace*{-2mm}
\caption{\label{fig7}
Data at the three finest lattice spacings together with a LO+NLO+NNLO fit (with
priors for the NNLO parameters). The breakup into LO/NLO/NNLO contributions
shows a convergence behavior, at the physical mass point, which is good for
$\fpi$ (top right) and even excellent for $\Mpi^2/m_{ud}$ (top left). For
higher quark masses the situation may get reversed. The respective values of
$\Lambda_3$ and $\Lambda_4$ are reasonably consistent with those from the pure
LO+NLO fit (bottom, blue bands copied from Fig.\,6). Figure taken from
Ref.\,\cite{Borsanyi:2012zv}.}
\end{figure}

With the restriction to the three finest lattice spacings ($a<0.13\fm$) the
data can even sustain a LO+NLO+NNLO joint chiral fit, provided we add (mild)
priors to stabilize the NNLO coefficients (which we are not interested in
anyway).
A typical behavior is shown in Fig.\,\ref{fig7}.
The point is that we can now perform a break-up into LO (green), LO+NLO (red)
and LO+NLO+NNLO (blue) part.
At the physical mass point the numerical values of $\fpi\,[\mr{MeV}]$ are
122.6, 130.7, 130.4, respectively, which I would term a ``good'' convergence
behavior (the first shift is by 6.6\%), and the numerical values of
$\Mpi^2/m_{ud}\cdot m_{ud}^\mr{phys}\,[10^2\MeV^2]$ are 186.3, 180.7, 181.4,
respectively, which I would call an ``excellent'' convergence behavior (the
first shift is by 3.0\%).
As the two lower panels of that figure indicate, this fit is not entirely
immune against changes of the chiral fit range (in particular if the lower
bound is increased), but the values of the NLO LECs $\Lambda_3, \Lambda_4$
stay reasonably consistent with what was obtained from the pure LO+NLO fit
(blue bands for $\mr{stat}$ and $\sqrt{\mr{stat}^2+\mr{syst}^2}$ errors).

We find $\bar\ell_3=3.16(10)_\mr{stat}(29)_\mr{syst}$ and
$\bar\ell_4=4.03(03)_\mr{stat}(16)_\mr{syst}$ besides the LECs at LO
\cite{Borsanyi:2012zv}.
A more extensive discussion of SU(2) LECs from the lattice along with some
world-averages is found in \cite{Colangelo:2010et}.
It turns out that to date there is no significant difference for a given SU(2)
LEC from $\Nf=2$ versus from $\Nf=2+1$ simulations.
Hence unquenching effects from $s$-loops seem to be mild.

\clearpage

%%%%%%%%%%%%%%%%%%%%%%%%%%%%%%%%%%%%%%%%%%%%%%%%%%%%%%%%%%%%%%%%%%%%%%%%%%%%%%%%

\section{Questions with the chiral SU(3) framework}

It is known from phenomenology that $m_s^\mr{phys}\simeq95\MeV$ (at $\mu=2\GeV$
in $\MSbar$ scheme) is at the edge of the regime where ChPT converges well.
The good news is that the lattice can vary $m_s$ around this value and explore
the issue in more detail.
The bad news is that many of the existing $\Nf=2+1$ studies are pounded with
additional convergence issues that come from $m_q^\mr{sea}\neq m_q^\mr{val}$.

\begin{figure}[tb]
\centering
\vspace*{-4mm}
\includegraphics[angle=-90,width=0.7\textwidth]{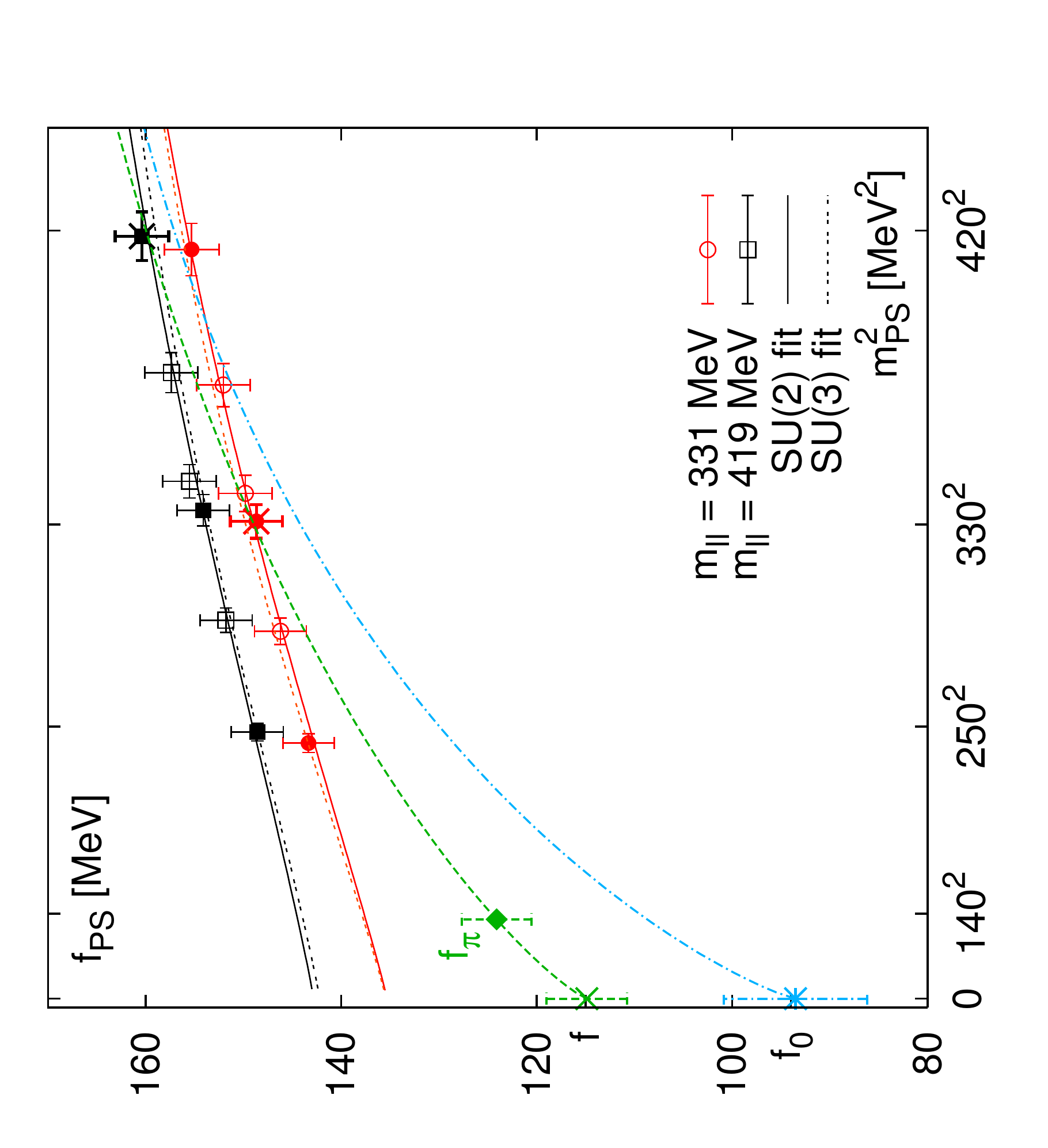}%
\vspace*{-4mm}
\caption{\label{fig8}
Partially quenched data by the RBC/UKQCD collaboration (as of 2008) for $\fpi$
(which they denote by $f_\mr{PS}$) versus $m_{ud}^\mr{val}$.
They feature two unitary data-points (bursts); the fits extract the decay
constants $f$ and $f_0$ in the SU(2) and SU(3) [unitary] chiral limits,
respectively. Figure taken from Ref.\,\cite{Allton:2008pn}.}
\end{figure}

An older paper worth discussing is Ref.\,\cite{Allton:2008pn}; their famous
plot is reproduced in Fig.\,\ref{fig8}.
It shows their partially quenched data on two ensembles (red and black) versus
$m_{ud}^\mr{val}$ and fitted with PQChPT.
This fit yields the unitary $\fpi$ in two theories:
($i$) as a function of $m_{ud}^\mr{sea}=m_{ud}^\mr{val}$ at fixed
$m_s^\mr{sea}=m_s^\mr{phys}$ [SU(2), green line] and ($ii$) as a function of
$m_{ud}^\mr{sea}=m_{ud}^\mr{val}=m_s^\mr{sea}=m_s^\mr{val}$ [SU(3), blue line].
I think three points should be emphasized.
First, the two unitary lines suggest $f/f_0\equiv F/F_0=1.2(1)$ which is
interesting because it specifies the amount of Zweig rule violation.
Second, as pointed out by the authors, the extrapolated values $f$ and $f_0$
lie significantly below the data.
Finally, one should keep in mind that the not-so-great convergence apparent in
this plot may --~at least in part~-- be due to the fact that it is unnatural
for PQChPT to accommodate nearly linear data (the curvature in the partially
quenched logs must be counterbalanced by higher-order terms).
In my opinion this calls for an investigation how the convergence pattern
depends on the width of the partially quenched direction.
For the progress achieved by RBC/UKQCD since publication of
Ref.\,\cite{Allton:2008pn} see \cite{Mawhinney}.

\begin{figure}[tb]
\centering
\includegraphics[width=0.485\textwidth]{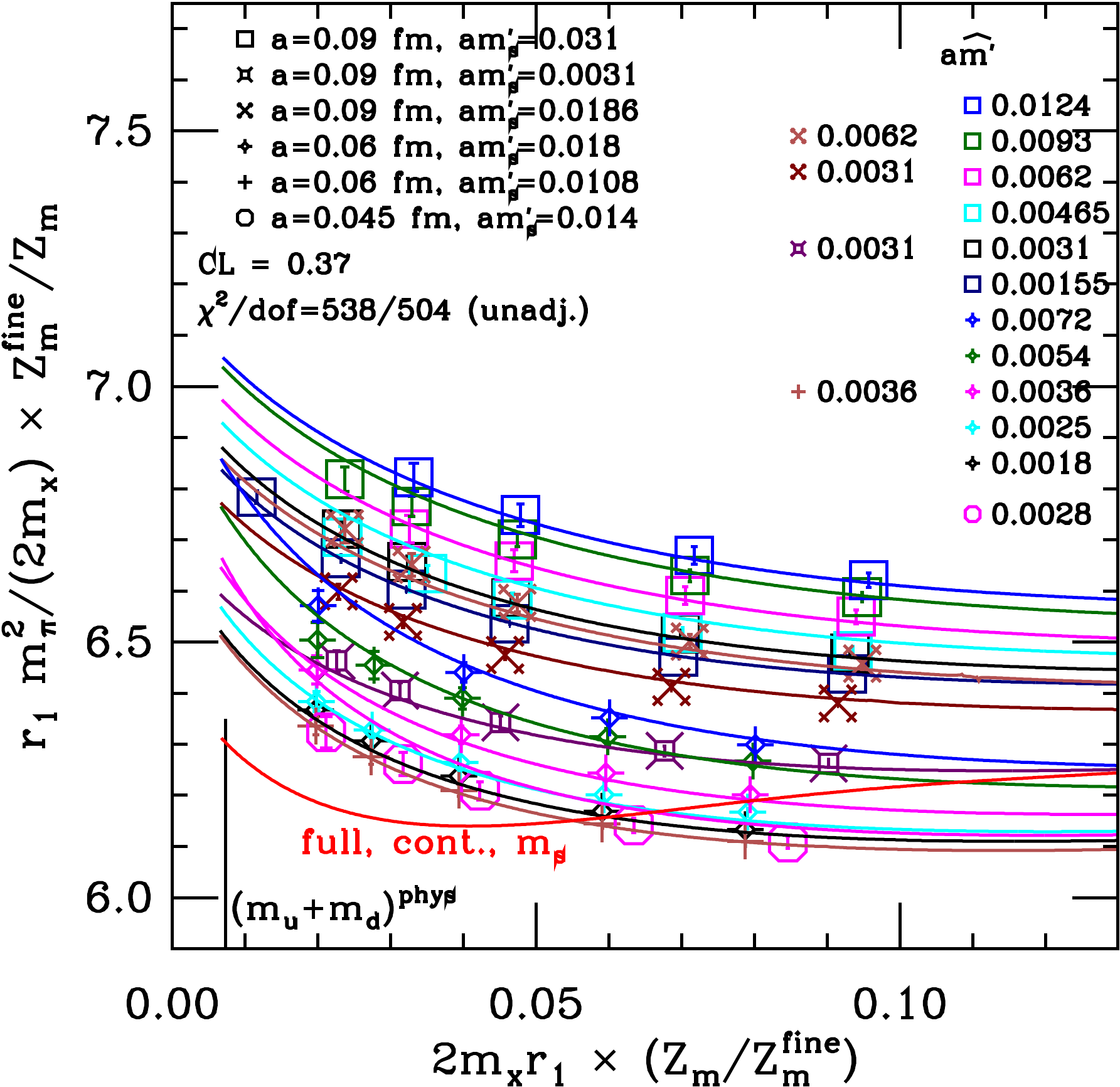}%
\includegraphics[width=0.515\textwidth]{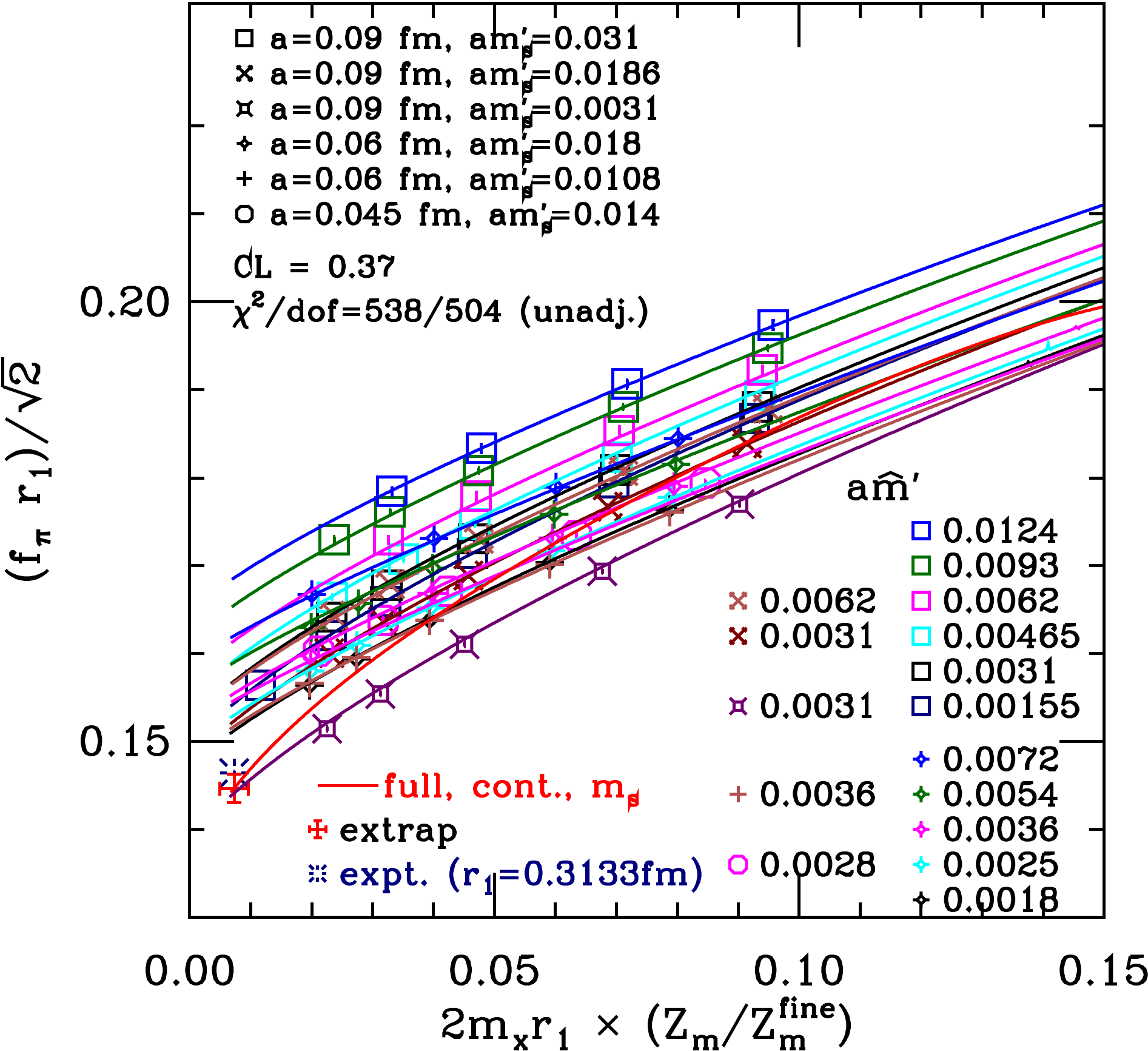}%
\caption{\label{fig9}
Partially quenched data by the MILC collaboration (as of 2010) for
$\Mpi^2/m_{ud}$ (left) and $\fpi$ (right) versus $m_{ud}^\mr{val}$. The
unitary continuum behavior at $m_s=m_s^\mr{phys}$ is shown in red.
Figure taken from Ref.\,\cite{Bazavov:2010hj}.}
\end{figure}

\begin{figure}[tb]
\centering
\includegraphics[width=0.490\textwidth]{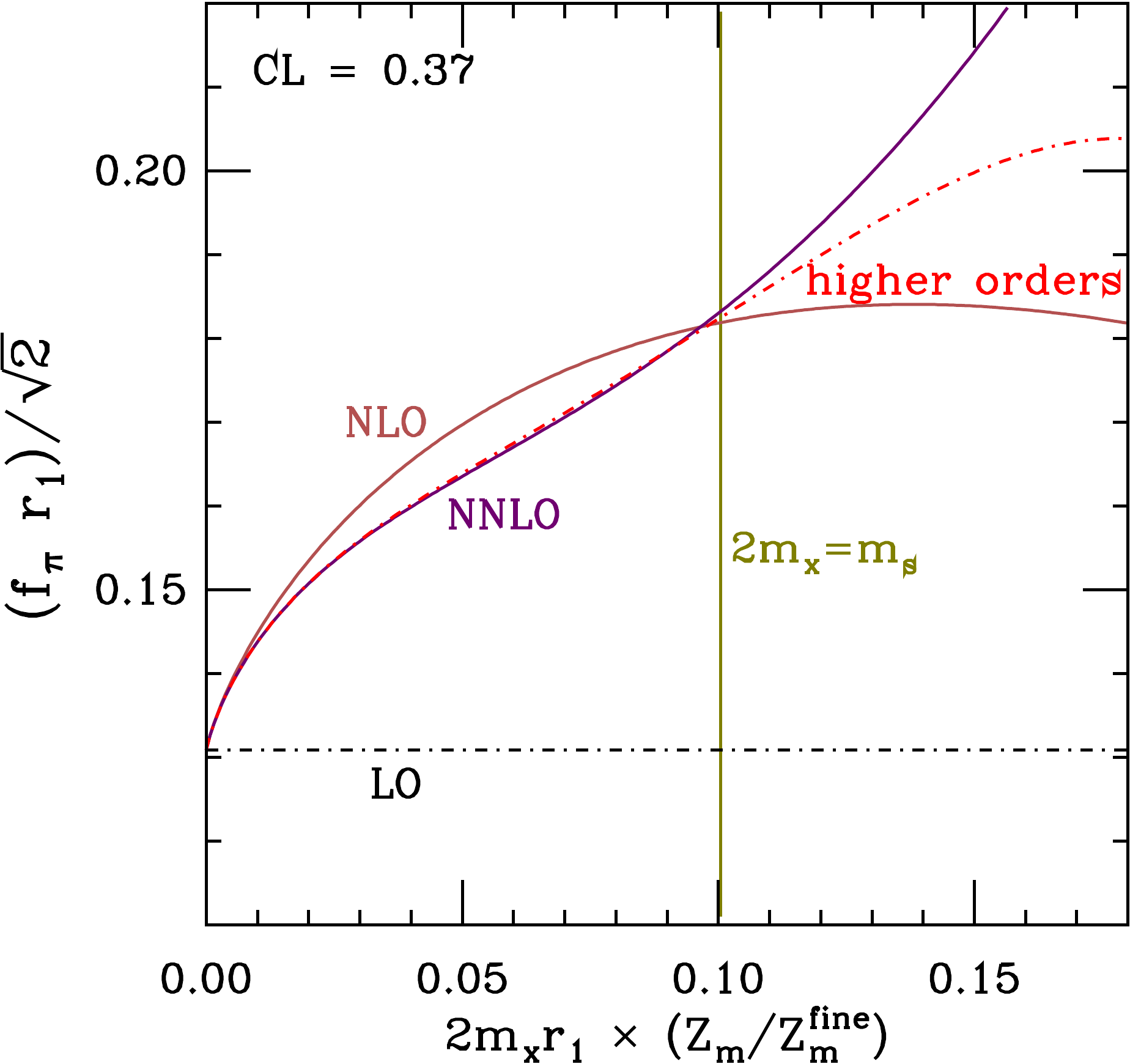}%
\includegraphics[width=0.505\textwidth]{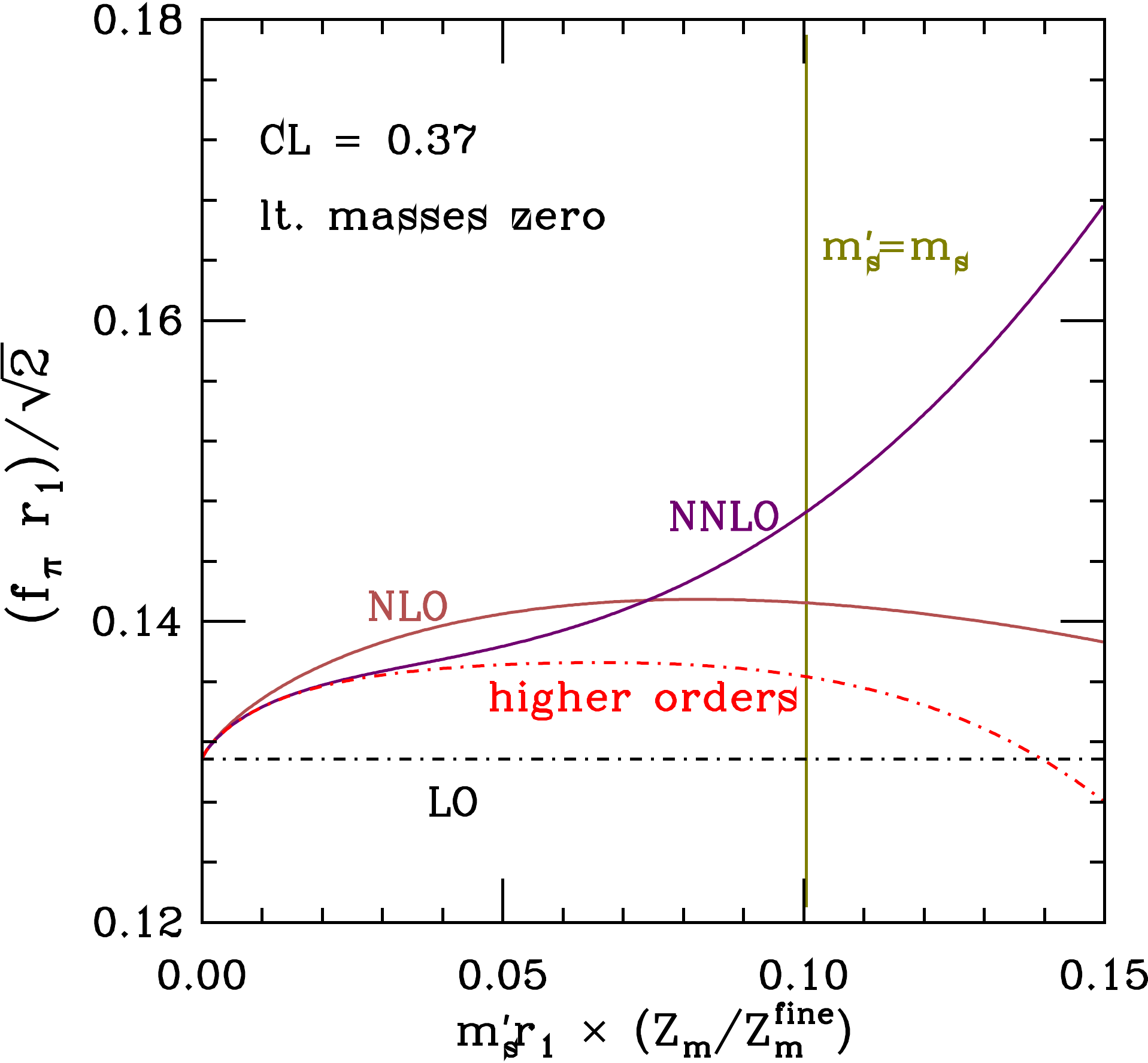}%
\caption{\label{fig10}
Breakup of the continuum extrapolated and unitary subset of the fit shown in
the right panel of Fig.\,9 into its LO/NLO/NNLO/higher-order-analytical contributions, both for
the $m_{ud}=m_s$ diagonal line in Fig.\,3 (left) and for the SU(2) chiral limit
as a function of $m_s$ (right). Figure taken from Ref.\,\cite{Bazavov:2010hj}.}
\end{figure}

Another collaboration with an interesting $\Nf=2+1$ dataset is MILC.
They have ensembles with $m_s\ll m_s^\mr{phys}$, i.e.\ additional green crosses
close to the $x$-axis in the cartoon of Fig.\,\ref{fig3}.
In Ref.\,\cite{Bazavov:2010hj} they display a fit to their full (partially
quenched) dataset along with the restriction of that fit to the unitary world
where $m_q^\mr{sea}=m_q^\mr{val}$ for both $q=ud$ and $q=s$ (the red ``full,
cont, $m_s$'' line in Fig.\,\ref{fig9}).
Fig.\,\ref{fig10} shows the breakup of this unitary restriction into
LO/NLO/NNLO/higher-order-analytical terms, both along the $m_{ud}=m_s$ diagonal
line in the cartoon (left panel) and along the $y$-axis of the cartoon as a
function of $m_s$ (right panel).
In the former case convergence seems to be good up to
$2m_{ud}\simeq m_s^\mr{phys}$ (marked by the green line).
In the latter case the SU(2) convergence seems to depend on $m_s$; specifically
near $m_s=m_s^\mr{phys}$ (labeled $m_s'=m_s$) the convergence seems rather poor.
This latter finding tends to be in conflict with the pattern observed in
Fig.\,\ref{fig7} from a direct SU(2) fit.

In short it seems fair to say that there are open issues regarding the
convergence of (extended versions of) SU(3) ChPT on $\Nf=2+1$ ensembles.
For numerical values of SU(3) LECs see \cite{Colangelo:2010et}.

\clearpage

%%%%%%%%%%%%%%%%%%%%%%%%%%%%%%%%%%%%%%%%%%%%%%%%%%%%%%%%%%%%%%%%%%%%%%%%%%%%%%%%

\section{Brief comment on the viability of $\;m_u=0$}

It is well known that ``\,$m_u=0$\,'' (in QCD) would provide a theoretically
appealing solution to the strong CP problem.
The question is just: Is it phenomenologically viable ?

\begin{figure}[tb]
\vspace*{-1mm}
\centering
\includegraphics[width=0.7\textwidth]{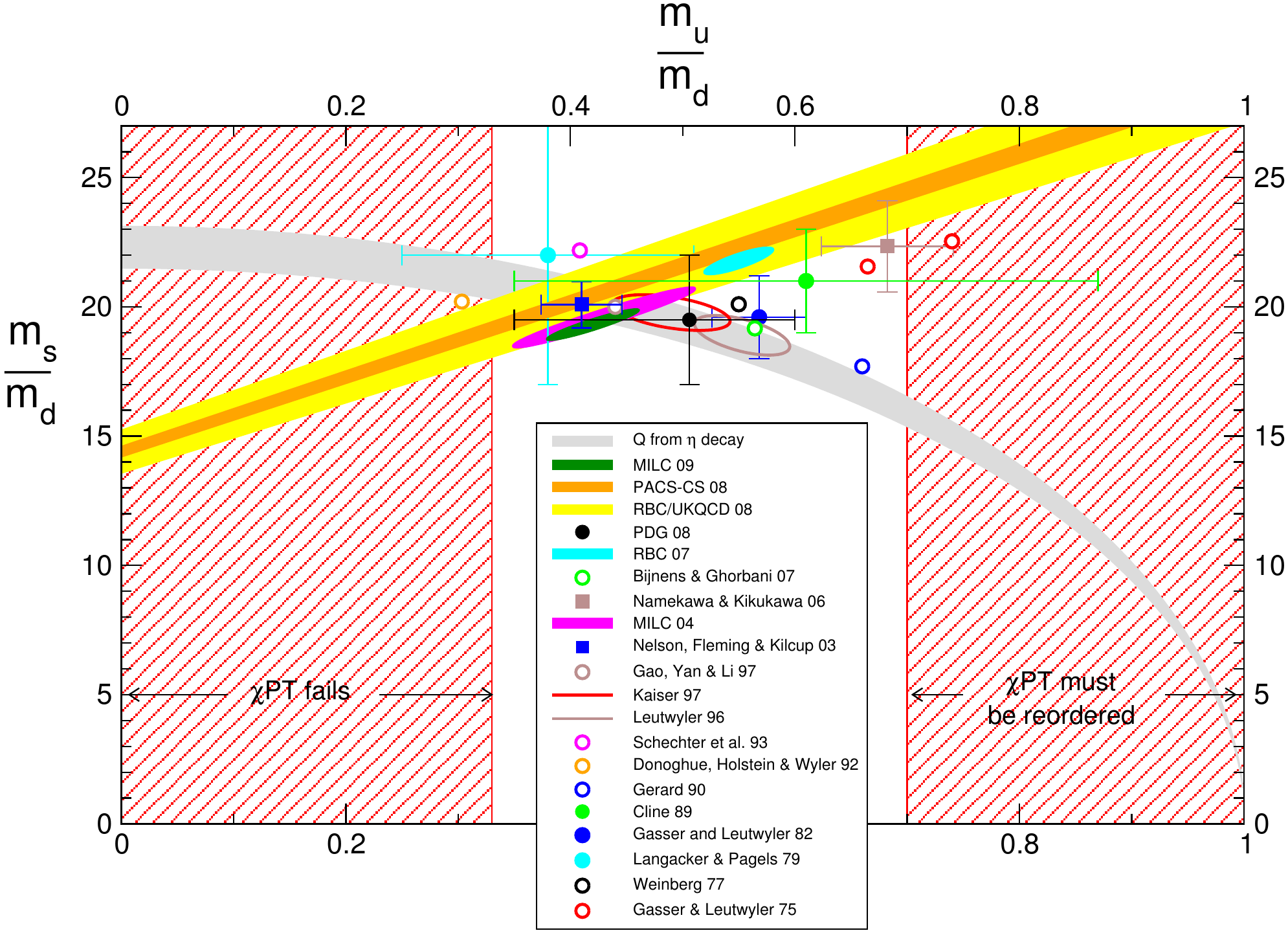}%
\caption{\label{fig11}
Some phenomenological and lattice determinations of $(m_u/m_d)^\mr{phys}$ and
$(m_s/m_d)^\mr{phys}$ along with shaded areas for which ChPT augmented for
electromagnetic effects would be in trouble if the experimental values of
$M_{\pi^+}^2-M_{\pi^0}^2$ and $M_{K^+}^2-M_{K^0}^2$ are untouched. Figure taken
from Ref.\,\cite{Leutwyler:2009jg}.}
\end{figure}

Fig.\,\ref{fig11} reproduces a plot from Ref.\,\cite{Leutwyler:2009jg}.
Leutwyler shows several results for $(m_s/m_d)^\mr{phys}$ versus
$(m_u/m_d)^\mr{phys}$ along with red bands which indicate that ChPT augmented
to account for electromagnetism fails to converge if the latter ratio would be
below $\sim\!0.3$ or above $\sim\!0.7$ (which apparently is not the case).
To avoid potential misunderstanding: There is no statement that ChPT+QED cannot
describe a world with an up/down quark mass ratio of, say, $0.1$ and
$\alpha_\mr{QED}\simeq1/137$, if $M_{\pi^+}^2-M_{\pi^0}^2$ and
$M_{K^+}^2-M_{K^0}^2$ change accordingly.
The statement is that this extended chiral framework fails to converge if the
meson mass splittings stay at their experimental values and nonetheless the
internal $m_u/m_d$ ratio is pinned to a value outside the white region.
In short the physics question is: Does this indicate that ``\,$m_u=0$\,'' is
phenomenologically not viable or does it, to the contrary, just signal an
inability of ChPT+QED to reconcile the beautiful solution with experimental
facts\,?

Over the years the lattice has made great progress at pinning down the quark
mass ratio $m_u/m_d$ (and also $m_s/m_{ud}$, both in QCD) independently, i.e.\
with steadily decreasing chiral input.
An early study by MILC used ChPT+QED in the pion/kaon system and found
$m_u/m_d=0.43(1)(8)$ \cite{Aubin:2004fs}.
A calculation by BMW used more robust information about strong isospin breaking
from $\eta\to3\pi$ decays and found $m_u/m_d=0.45(1)(3)$ \cite{Durr:2010vn}.
There are several new results with quenched/full QED on full QCD backgrounds,
e.g.\ Blum et al.\,\cite{Blum:2010ym}, PACS-CS \cite{Aoki:2012st}, RM123
\cite{deDivitiis:2013xla} and BMW \cite{Portelli}, which find significant but
non-dramatic corrections to Dashen's theorem, indicating that $m_u/m_d$ is away
from zero by $O(10)$ standard deviations and well inside the white region in
Fig.\,\ref{fig11}.

Of course, one may choose to wait for a fullQCD+fullQED study (without
reweighting), but with hindsight one may say that nature solves the strong CP
problem not by ``\,$m_u=0$\,''.

\clearpage

%%%%%%%%%%%%%%%%%%%%%%%%%%%%%%%%%%%%%%%%%%%%%%%%%%%%%%%%%%%%%%%%%%%%%%%%%%%%%%%%

\section{Summary}

Let me summarize the salient points in a few short statements:
\begin{enumerate}
\itemsep-2pt
\item
The lattice community is at the point where physical quark masses can be
simulated, i.e.\ ensembles with physical values of $(\Mpi^2,2\Mka^2-\Mpi^2)$
in large enough boxes and at several lattice spacings can be generated.
As a result chiral extrapolation formulas are now less important (while finite
volume correction formulas are still in high demand), and the lattice is in a
unique position to compute the chiral LECs from first principles.
\item
The SU(2) framework is best served by current $\Nf=2$ and $\Nf=2+1$ simulations
where (in the latter case) $m_s\simeq m_s^\mr{phys}$.
For $m_{ud}\simeq m_{ud}^\mr{phys}$ the ChPT convergence seems to be rapid.
The SU(2) LECs from $\Nf=2$ and $\Nf=2+1$ simulations are logically different,
but currently no numerical difference is seen, i.e.\ unquenching effects due to
$s$-loops seem to be mild.
\item
The SU(3) framework requires data with $m_s\ll m_s^\mr{phys}$ to control ChPT
systematics, as shown by MILC.
There are issues regarding the convergence pattern as well as the size of
unitarity violations and/or cut-off effects that can be parameterized by
extended versions of ChPT.
\item
Given the experimental values of the meson mass splittings
$M_{\pi^+}^2-M_{\pi^0}^2$ and $M_{K^+}^2-M_{K^0}^2$, the chiral framework with
electromagnetic effects would be in trouble if $m_u/m_d$ (in QCD) would be
significantly different from a value $\sim\!0.5$.
Evidence is mounting that this is not a deficiency of ChPT -- there is a number
of lattice results which exclude the esthetically pleasing solution
``\,$m_u=0$\,'' to the strong CP problem at the multi-sigma level.
\end{enumerate}

\smallskip\smallskip\smallskip

\noindent
{\bf Acknowledgments}:
 I would like to thank my colleagues in the Budapest-Marseille-Wuppertal
 collaboration on general grounds and     Enno Scholz from Regensburg for a very
 enjoyable collaboration which led to Ref.\,\cite{Borsanyi:2012zv}.
%Thanks go to          my colleagues in the Budapest-Marseille-Wuppertal
%collaboration on general grounds and to Enno Scholz from Regensburg for a very
%enjoyable collaboration which led to Ref.\,\cite{Borsanyi:2012zv}.
This work is partly supported by the German DFG through SFB-TR-55.

%%%%%%%%%%%%%%%%%%%%%%%%%%%%%%%%%%%%%%%%%%%%%%%%%%%%%%%%%%%%%%%%%%%%%%%%%%%%%%%%

\end{document}